\documentclass[referee]{aa} 		
\usepackage{epsfig,graphicx,times}
\newcommand{\kms}{{{km~s}$^{-1}$}}
\newcommand{\teff}{{$T_\mathrm{eff}$}}

\newcommand{\vt}{{$v_\mathrm{t}$}}

\begin{document}
\title{Chemical compositions of Four B-type Supergiants in the SMC Wing}

\subtitle{}

\author{J.-K. Lee\inst{1}
\and W.R.J. Rolleston\inst{1}
\and P.L. Dufton\inst{1}
\and R.S.I. Ryan\inst{1}
} 

\institute{Department of Pure \& Applied Physics, The Queen's University 
           of Belfast, BT7 1NN, Northern Ireland, UK       
}
\offprints{J.-K. Lee, \email{J.K.Lee@qub.ac.uk}}

\date{Received May 2004 / Accepted September 2004}

\abstract{High-resolution UCLES/AAT spectra of four B-type supergiants in the
SMC  South East Wing have been analysed using non-LTE model atmosphere
techniques to determine their atmospheric parameters and chemical compositions.
The principle aim of this analysis was to determine whether the very low metal
abundances ($-$1.1 dex compared with Galactic value) previously found in the
Magellanic Inter Cloud region (ICR) were also present in SMC Wing. The chemical
compositions of the four targets are similar to those found in other SMC
objects and appear to be incompatible with those deduced previously for the
ICR. Given  the close proximity of the Wing to the ICR, this is difficult to
understand and some possible explanations are briefly discussed.


}

\titlerunning{B-type Supergiants in the SMC Wing}

\maketitle

\section{Introduction}

\begin{figure}
\hspace*{-4mm}\psfig{file=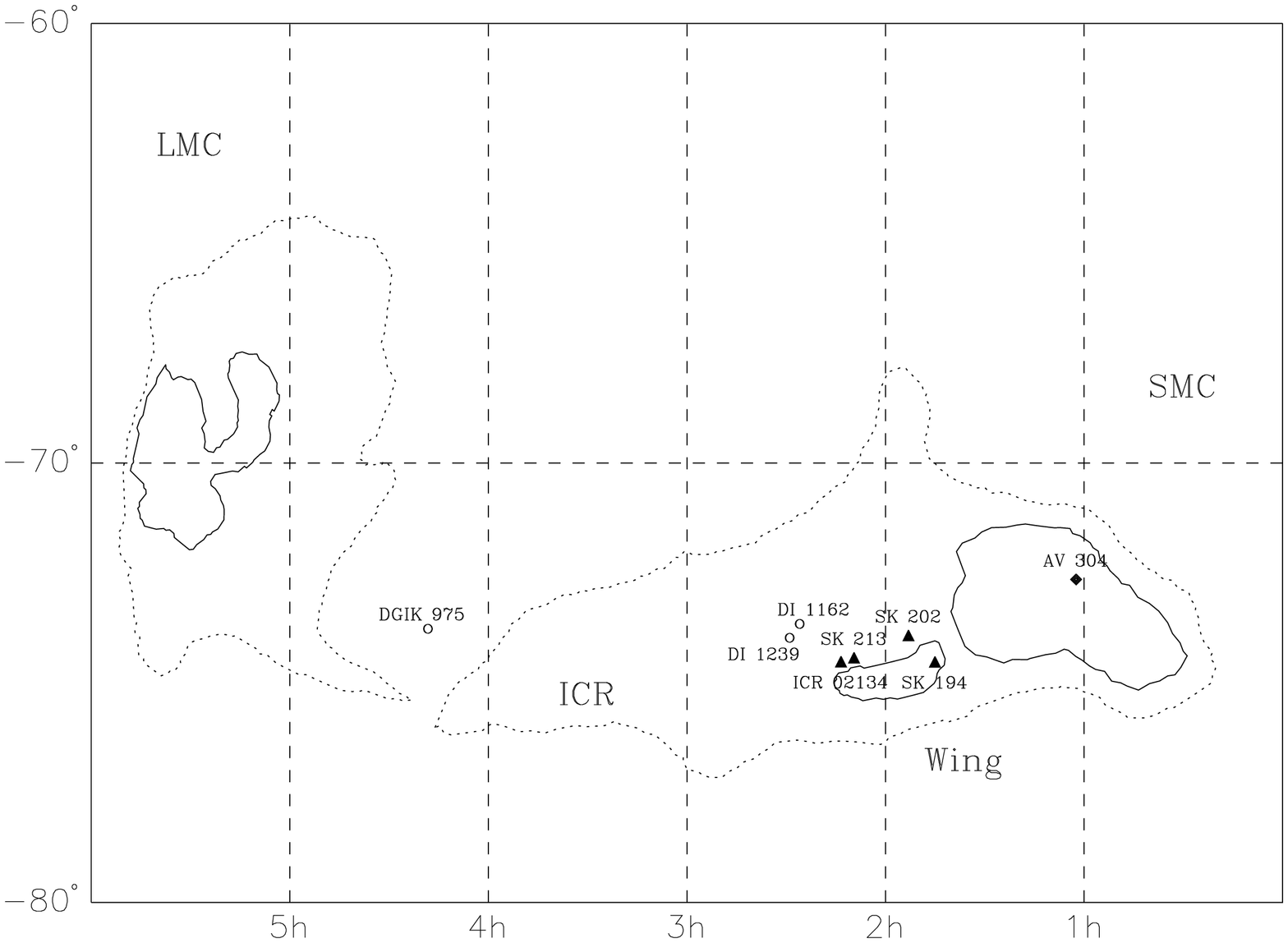,angle=90,width=9cm}
\caption{Schematic diagram of the LMC, SMC and the Inter-Cloud Region. Solid
lines define the stellar concentrations within the Magellanic Clouds, while the
dashed lines show the H~{\sc i} envelopes of the Magellanic system. The
positions of our targets within the SMC wing are marked by solid black
triangles. Other targets have been discussed by Rolleston et al.\
(\cite{Rol99}, circles;  \cite{Rol03},  diamond). } 
\label{smcw_fig2}
\end{figure}

Between the Large (LMC) and Small (SMC) Magellanic Clouds, two satellite
galaxies of our own Milky Way, lies a  column of gas  known as the Magellanic
Bridge. This Bridge greatly exceeds the galaxies' tidal limit, and is widely
considered to be a remnant of the gravitational interactions between the
Magellanic Clouds and the Milky Way. Numerical simulations by, for example,
Gardiner \& Noguchi (\cite{Gar96}) and  Sawa, Fujimoto \& Kumai (\cite{Saw99})
imply that the tidal effects from a close encounter of the two Magellanic
Clouds about $\sim$200 Myr ago produced a tidal bridge and tail structure and
subsequently triggered an episode of star formation.  This tail/bridge
structure  is believed to constitute the inter-cloud region (ICR)  and would be
seen as overlapping in the sky, with  the tail material modeled to be more
distant than the bridge and with a  larger radial velocity (Gardiner \& Noguchi
\cite{Gar96}). The SMC wing is located to the east of the SMC and enveloped by
neutral hydrogen gas (see Fig.~\ref{smcw_fig2}).  In optical images it appears
as a large cloud of faint stars extending 6.5\degr\/ eastward from the SMC
toward the LMC, and contains young stellar component (e.g.\ Westerlund \&
Glaspey \cite{Wes71}; Kunkel \cite{Kun80}) that indicates recent star formation
history. According to the predictions of the models (Kunkel et
al.\,\cite{Kun94}), this Wing constitutes the densest regions of the
bridge/tail material along the line-of-sight.

The tail/bridge material, if it is indeed the product of the past encounter(s)
of the Clouds, would be expected to have the metallicity of its progenitors.
However, a recent study by Rolleston et al.\ (\cite{Rol99}) of 3 B-type dwarfs
in the ICR reported that they had a metallicity lower by $-$1.1 dex than that
of our Galaxy, and this is also significantly lower than  the values found for
the SMC and LMC (e.g.\, Korn et al.\ \cite{Korn}; Kurt et al.\ \cite{Kur99};
Rolleston et al.\ \cite{Rol93, Rol96, Rol03};  Venn \cite{Ven99}). A possible
explanation is that the inter-cloud  material could have been stripped off from
either of the galaxies at a  much earlier encounter (about $\sim$8 Gyr ago;
Kobulnicky \& Skillman \cite{Kob97}; Da Costa \& Hatzidimitriou \cite{DaC98}),
when less nucleo-synthetic processing of interstellar medium (ISM) of the SMC
had occurred. However, this would be inconsistent with the numerical
simulations  discussed above that imply  this ICR was formed relatively
recently and also with the presence of early-type stars. 

Here we attempt to confirm and extend the results of Rolleston et al.\ by
investigating the chemical composition of the SMC wing. Using intermediate 
dispersion spectroscopy, we have identified suitable targets and here present
high-resolution spectroscopy of four of these objects. By investigating the
metallicity of these stars, we can address questions concerning the chemical
homogeneity, or inhomogeneity, of the two distinct components, as well as
possibly the tidal origin of the SMC wing material.   

\begin{table*}
\caption{Observational summary of the four program stars}\label{tab:obs}
\begin{tabular}{lccrrrrrcc}
\hline\hline\noalign{\smallskip}
 Star           &       RA,  Dec.         &  V   &  AAT/   & $T_{\mathrm exp}$& $v_{\mathrm lsr}$ & S/N\\
                &       (2000.0)          & [mag]& UCLES   & [sec]            & [\kms] & \\ 
\noalign{\smallskip}
\hline\noalign{\smallskip}
ICR\,02134-7431 &  02 13 26, $-$74 31 20  & 12.3 & 1999-10-25 &  7\,500 & +197&  80 \\  
SK\,194         &  01 45 03, $-$74 31 32  & 11.7 & 2000-10-14 &  6\,600 & +214& 140 \\
SK\,202         &  01 53 03, $-$73 55 30  & 12.3 & 2000-10-14 & 25\,200 & +180& 125 \\
SK\,213         &  02 09 30, $-$74 26 00  & 13.7 & 2000-10-16 &  7\,200 & +203&  60 \\
\noalign{\smallskip}\hline
\end{tabular}\\
\end{table*}

\begin{figure}
\hspace*{-4mm}\psfig{file=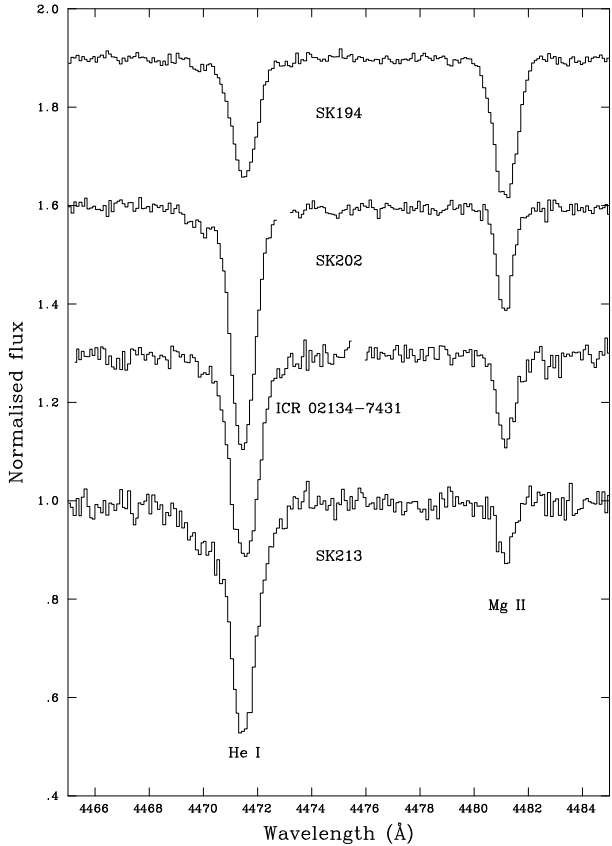,width=9cm}
\caption{Normalised spectra of the 4 targets covering the region 4465 to
4485~\AA, containing a \ion{He}{i} and a \ion{Mg}{ii} line. The spectra were
rebinned to a pixel size of 0.1~\AA. In SK\,213 the \ion{Mg}{ii} line has 
an equivalent width of approximately 70~m\AA.} \label{smcw_fig1}
\end{figure}

\section{Observations and data reduction}
\label{obs}

The high-resolution spectroscopic data presented here were obtained during two
observing runs with the 3.9-m Anglo-Australian Telescope (AAT) in October 1999
and 2000. The University College of London \'{E}chelle Spectrograph (UCLES) was
used with the 31 lines mm$^{-1}$ grating and with a TeK 1K$\times$1K CCD,
providing complete spectral coverage between $\lambda\lambda$3900--4900~\AA\/
at a FWHM resolution of $\sim$0.1~\AA. Stellar exposures were divided into
either 1\,200 or 1\,500 sec segments with the total observing times being
listed in  Table~\ref{tab:obs}. These exposure times should be treated with
some caution as the observing conditions varied significantly during the two
observing runs. Stellar exposures were bracketed with Cu-Ar arc exposures for
wavelength calibration. 

The two dimensional CCD datasets were reduced using standard procedures within
{\sc iraf} (Tody \cite{Tody}). Preliminary processing of the CCD frames such as
over-scan correction, trimming of the data section and flat-fielding were
performed using the {\sc ccdred} package (Massey \cite{Mas97}), whilst
cosmic-ray removal, extraction of the stellar spectra, sky-subtraction and
wavelength calibration were carried out using the {\sc specred} (Massey et al.\
\cite{Mas92}) and {\sc doecslit} (Willmarth \& Barnes \cite{Wil94}) packages. 
The spectra were then co-added and re-binned to 0.1~\AA, which was adequate for
the relatively broad metal absorption lines (with typical
full-width-half-maximum widths of 1--2~\AA) found in these supergiant spectra.
The signal-to-noise (S/N) ratios found in the continuum are summarized in
Table~\ref{tab:obs} for a wavelength of approximately 4300~\AA. Note that care
has been taken to include different parts of the blaze profile in these
measurements so that they should be representative of the overall quality of
the spectra. In Fig.~\ref{smcw_fig1}, we illustrate the quality of the spectra
for the region 4465--4485~\AA, which was chosen as it contains lines of
\ion{He}{i} and \ion{Mg}{ii} observable in all four targets. For the hottest
star, SK\,213, the \ion{Mg}{ii} line is relatively weak, with an equivalent
width of approximately 70~m\AA\, but is still clearly visible. Indeed lines
with equivalent widths greater than 40~m\AA\ could be measured in all our
targets with weaker features being identified in the two targets with the
highest S/N ratios.

Heliocentric stellar radial velocities (summarised in Table~\ref{tab:obs}) and 
equivalent widths (EWs) of lines were measured using the {\sc starlink}
spectrum  analysis program {\sc dipso} (Howarth et al.\ \cite{How94}). Low
order polynomials were  used to represent the adjacent  continuum regions for
normalisation and then  Gaussian profiles were then fitted to the metal and
non-diffuse helium lines  using non-linear least square routines. As discussed
by Ryans et al.  (\cite{Rya03}), the profiles of metal absorption lines in the
spectra of  early-type supergiants are well represented by a Gaussian profile
but tests  using a different profile shape showed that this assumption was not
critical.  For hydrogen lines, the above method was not feasible as there was
insufficient continuum regions for normalisation. Instead the continua of the
adjacent  orders were moved to a common wavelength scale, fitted by low order 
polynomials, which were then merged. This estimate of the blaze profile  was
then used to rectify the order containing the hydrogen lines. Profiles  were
then extracted with the continuum levels being defined at $\pm$16~\AA\/  from
the line centre.

\section{Model Atmospheres}\label{model}

\subsection{Model atmosphere calculations} 

The analysis is based on grids of non-LTE model atmosphere calculated using
the  codes {\sc tlusty} and {\sc synspec} (Hubeny \cite{Hub88}; Hubeny \& Lanz
\cite{Hub95}; Hubeny et  al.\ \cite{Hub98}).  Details of the methods can be
found in Ryans et al.\ (\cite{Rya03}), while the grids will be discussed in
more detail by Dufton et  al.\ (\cite{Duf04}). 

Briefly four grids are being generated with base metallicities corresponding 
to our galaxy ($[\frac{Fe}{H}]$ = 7.5 dex) and with metallicities reduced by
0.3,  0.6 and 1.1 dex. These lower metallicities were chosen so as to be
representative  of the LMC, SMC and ICR material. For each base metallicity,
approximately 3\,000 models have been calculated covering a range of effective
temperature from 12\,000 to 35\,000~K, logarithmic gravities (in cm s$^{-2}$)
from 4.5 dex down to close to the  Eddington limit (which will depend on the
effective temperature) and microturbulences  of 0, 5, 10, 20 and 30 km
s$^{-1}$. Then for any set of atmospheric parameters,  five models were
calculated keeping the iron abundance fixed but allowing the  light element
(e.g. C, N, O, Mg, Si) abundances to vary from +0.8 dex to $-$0.8 dex around
their base values. Effectively this approaches assumes that the line 
blanketing and atmospheric structure is dominated by iron and hence that
the light element abundances can be varied without significantly affecting
this  structure. Tests discussed in Dufton et al.\ (\cite{Duf04}) appear to
confirm that this approach is reasonable. 

These models are then used to calculate spectra, which in turn provide
theoretical  hydrogen and helium line profiles and equivalent widths for light
metals for a  range of abundances. Note that as we keep the iron abundance
fixed within any  given grid, we have less extensive data for this element. The
theoretical equivalent  widths are then available via a GUI interface written
in IDL, which allows the user  to interpolate in order to calculate equivalent
widths and/or abundance estimates  for approximately 200 metal lines for any
given set of atmospheric parameters. Ryans et al.\ (\cite{Rya03}) reported that
the increments of 0.4 dex used in our grids were fine  enough to ensure that no
significant errors were introduced by the interpolation procedures. Full
theoretical spectra are also available for any given model. Details of the
model grids, atomic data used in the line strength calculations and  wavelength
ranges used in the equivalent width calculations are available at 
http://star.pst.qub.ac.uk/. In summary these grids allow a user-friendly
non-LTE analysis of hydrogen and helium  line profiles and of the profiles and
equivalent width of the lines of light elements over a range of iron line
blanketing and atmospheric parameters appropriate  to B-type stars in our
Galaxy and in Local Group galaxies such as the Magellanic Clouds.

During the analysis, one of the program stars, SK\,194, was found  to have
atmospheric parameters below our lowest grid points in both effective 
temperature and surface gravity. We have therefore calculated additional models
for \teff\/ = 11\,000~K and log\,$g$ = 1.6 to enable the spectrum of this  star
to be analysed.

\begin{table}  

\caption{Atmospheric parameters of the program stars derived for  two 
microturbulences (\vt), 10 and 20 \kms, using the SMC and ICR model
grids. Each pair of effective temperature (in K) and logarithmic gravity (in
cm~s$^{-2}$) is written as numbers separated by a comma.}\label{tab:atm_stars}

\begin{tabular}{rcccc}
\hline\hline\noalign{\smallskip}
   \vt      & IC\,02134-7431  & SK\,202    & SK\,213    & SK\,194 \\
\noalign{\smallskip}   \hline\noalign{\smallskip}  
 & \multicolumn{4}{l}{SMC grid}    \\
   10 & 15\,000, 2.1 & 14\,750, 2.2 & 20\,000, 2.9 & 11\,700, 1.7 \\
   20 & 14\,750, 2.1 & 14\,750, 2.2 & 20\,500, 2.9 & 11\,500, 1.7 \\
 & \multicolumn{4}{l}{ICR grid}    \\
   10 & 15\,000, 2.1 & 14\,750, 2.2 & 21\,000, 2.9 & 12\,000, 1.7 \\
   20 & 14\,750, 2.1 & 14\,750, 2.2 & 20\,750, 2.9 & 11\,500, 1.7 \\
\noalign{\smallskip} \hline     
\end{tabular}
\end{table}    

\subsection{Estimation of atmospheric parameters}

The estimation of atmospheric parameters involves an iterative process as they
are interrelated.  Firstly adopting appropriate values of the surface gravity
(using the calibration of McErlean et al.\ \cite{McE99}), the effective
temperature was estimated using the ionisation balances due to silicon. For the
coolest star, SK\,194, only one ionization level of silicon was observed, in
this case the He\,{\sc i} line profiles were used to estimate the effective
temperature. Although the helium lines are very  sensitive to effective
temperature in this regime, this method  implicitly assumes that the helium
abundance of SK\,194 is normal.  Using these preliminary effective temperature
estimates, the surface gravities were  then estimated by comparing the observed
hydrogen Balmer line profiles with theoretical calculations.  These new gravity
estimates were used as the starting point in the next iteration to estimate
effective temperatures, and the process was  repeated until convergence was
obtained. The entire procedure was undertaken for both \vt\/ = 10 and 20 \kms,
consistent with the  microturbulences found for supergiants by Gies and Lambert
(\cite{Gie92}), McErlean et al.\ (\cite{McE99}) and Trundle et al.\
(\cite{Tru04}). 

As discussed above, the ionisation balance of silicon lines was used for 
ICR\,02134-7431 and SK\,202 which show both \ion{Si}{ii} and {\sc iii} lines.  
For SK\,213 whose observed spectra exhibited only \ion{Si}{iii} lines, the 
absence of Si\,{\sc ii} and Si\,{\sc iv} lines was respectively used to provide
constraints on the lower and upper limit to the effective temperature.  The
mean of the derived lower and upper limits  (which differed by $<$ 1\,000~K)
was taken to be the effective temperature of the star. The derived parameters
are summarised in Table~\ref{tab:atm_stars}, using the grids with an iron
abundance of 6.9 dex (SMC grid) and 6.4 dex (ICR grid).

\subsection{Photospheric Abundances}  
\label{abund}

The adopted atmospheric parameters (listed in Table~\ref{tab:atm_stars})  were
used to derive absolute non-LTE abundances from the observed equivalent widths
for the programme stars. The grids with iron abundance appropriate  to either
the SMC or ICR were used (together with the appropriate atmospheric parameters)
but as the differences in the abundance estimates were small  only the results
for the SMC grid are presented in Table~\ref{tab:abs}.

\begin{table*}\caption{Absolute abundances of the program stars in the {\sc
tlusty} SMC grid for two microturbulences, 10 and 20 \kms. The last three
columns present results of three other recent studies of SMC stars (see
text).}\label{tab:abs}

\begin{tabular}{rrlrlrlrlccc}
\hline\hline\noalign{\smallskip}
            & \multicolumn{2}{c}{ICR\,02134-7431} & \multicolumn{2}{c}{SK\,202} 
	    & \multicolumn{2}{c}{SK\,213}    & \multicolumn{2}{c}{SK\,194} 
	    & \multicolumn{2}{c}{SMC supergiants}  & AV\,304 \\
	       & VT10& VT20    & VT10& VT20    & VT10& VT20    & VT10& VT20 
	       & TLPD   & DRTLHL  \\
\noalign{\smallskip}\hline\noalign{\smallskip}   
C\,{\sc ii}        & 6.7 & 6.7 (1) & 6.8 & 6.8 (1) & 7.2 & 7.2 (2) &  --~ & \,~-- 
& 7.30 & 7.06 & 7.36  \\
N\,{\sc ii}        & 7.4 & 7.3 (1) & 7.6 & 7.4 (2) & 7.7 & 7.6 (9) & 7.9 & 8.0(1)
& 7.67 & 7.42 & 6.55 \\
O\,{\sc ii}        &  --~ & \,~--  & --~ & \,~--   & 8.3 & 8.0 (15)&  --~ & \,~--  
& 8.15 & 8.09 & 8.13 \\
Mg\,{\sc ii}       & 6.7 & 6.6 (1) & 6.6 & 6.5 (1) & 6.7 & 6.7 (1) & 6.8 & 6.6(1)
& 6.78 & 6.79 & 6.77 \\
Si\,{\sc ii/iii}   & 6.7 & 6.6 (4) & 6.9 & 6.8 (4) & 7.2 & 6.7 (3) & 7.1 & 6.8(2)
& 6.74 & 6.85 & 6.76  \\
Fe\,{\sc ii}   &  --~ & \,~--  &  --~ & \,~--  &  --~ & \,~--  & 6.2 & 6.1(15)
& -- & -- & -- \\
\hline\noalign{\smallskip}      
\end{tabular}
\end{table*}

\section{Discussion}       

The absolute abundances of metallic lines are summarised in Table~\ref{tab:abs}
on a logarithmic scale with [H] $\equiv$ 12.0 dex. An observational uncertainty
of $\pm$10~m\AA\/ in the equivalent width estimate of an individual line would
change an abundance estimate by typically $\pm$0.15 dex. Where several lines
due to an ionic species are observed the corresponding error in the mean
abundance would be expected to be smaller. The abundance estimates may also be
affected by errors in the adopted atmospheric parameters. The results,
presented in Table~\ref{tab:abs} for microturbulences of \vt\/ = 10 and 20
\kms, lead to changes in the abundance estimates of typically $\pm$0.1 dex,
with larger uncertainties for in a few cases. For the effective temperature and
gravity, numerical tests adopting error estimates of $\pm$1\,000~K and $\pm$
0.2\,dex respectively, lead to changes in the abundance estimates of typically
0.1 to 0.3 dex.  Finally there will be uncertainties due to the intrinsic
assumptions made in the model atmosphere calculations and in particular the
neglect of the stellar wind. However, Trundle et al.\  (\cite{Tru04}) have used
the non-LTE unified model atmosphere code {\sc fastwind} first introduced by
Santolaya-Rey, Puls \& Herrero (\cite{San97}) to analyse the spectra of SMC
supergiants. For two targets, these results have been compared with those
deduced using {\sc tlusty} (see Dufton et al.\  \cite{Duf04}) and the
comparison indicates that the neglect of the wind may not be a serious source
of error.  Hence we conclude that an error estimate of $\pm$0.3 dex might be
appropriate to our abundance estimates.

The main purpose of this analysis is to investigate whether the very low metal
abundances for the ICR found by Rolleston et al.\ (\cite{Rol99}) are also
present in our SMC wing targets. To provide a comparison for our results, we
have utilised three other recent studies of hot stars in the SMC. Trundle et
al.\  (\cite{Tru04}, TLPD) studied 7 supergiants using the non-LTE unified
model atmosphere code {\sc fastwind}. Dufton et al.\ (\cite{Duf04}) have
undertaken analysis of 9 supergiants using the same non-LTE grids as adopted
here. As discussed above,  the two analyses yield similar results for two
targets in common indicating that both TLPD and Dufton et al.\ should form
suitable comparisons. Lennon et al.\ (\cite{Len03}) corrected the LTE analysis
of AV\,304, a sharp lined B0V star, of Rolleston et al.\ (\cite{Rol03}) in
order to allow for non-LTE effects. Although these corrections can  only be
approximate, they would appear to provide estimates that agree well with a
full non-LTE analysis currently being undertaken by the authors. The results of
all these analyses are summarized in Table~\ref{tab:abs}, the first two
being for stars at a similar evolutionary to our targets, whilst the third
should reflect the current chemical composition of the SMC interstellar medium.
The three analyses yield different N (and to a lesser extent O)
abundances estimates for reasons that are discussed below. However the
excellent agreement for the O, Mg and Si abundances provide indirect support
for their use as comparators for our SMC wing targets. Below we briefly discuss
the estimates obtained for each element.

{\bf Carbon:} The carbon abundances in the Wing targets appear to be lower for
ICR\,02134-7431 and SK\,202 than for other SMC targets. The value for SK\,213
appear relatively normal, whilst no \ion{C}{ii} features were measurable in
SK\,194. However these results must be treated with caution for the estimates
have used the doublet at 4267~\AA. As discussed by, for example, Sigut
(\cite{Sigut}), this element suffers from significant non-LTE effects which are
difficult to model. Additionally, Lennon et al.\ (\cite{Len03}) found that,
even in a non-LTE analysis, this doublet appears to yield lower abundance
estimates (by 0.3--0.4 dex) than is found from other \ion{C}{ii} lines. Indeed
the carbon abundance for AV\,304 presented in Table~\ref{tab:abs} has been
corrected for this effect and this may explain the difference between the C
abundance estimates in Dufton et al.\ (\cite{Duf04}) and for AV\,304. Hence we
do not believe that there is any significant evidence for the carbon abundance
of our Wing targets being lower than that for the SMC in general.

{\bf Nitrogen:} Nitrogen abundances were found for all four targets and range
from 7.3 to 8.0 dex. TLPD also found a significant variation in their abundance
estimates that they ascribed to different degrees of contamination by
nucleo-synthetically processed material. This is also consistent with the far
larger nitrogen abundances found in the supergiant samples than in AV\,304 or
in analysis of SMC \ion{H}{ii} regions (see, for example, Kurt et al.\
\cite{Kur99}). Hence the N abundance estimates for our targets appear to be
compatible with those for other evolved SMC supergiants.

{\bf Oxygen:} The \ion{O}{ii} spectrum is only observed in our hottest target
with the abundance estimates being sensitive to the adopted microturbulence.
However they are consistent with those found in other SMC B-type stars.

{\bf Magnesium:} All the abundance estimates are based on the \ion{Mg}{ii}
doublet at 4481~\AA. However they show little variation and are in excellent
agreement with the other analyses.

{\bf Silicon:} These abundance estimates are based on both \ion{Si}{ii} and
\ion{Si}{iii} features. However, as the silicon ionization equilibrium has been
used to constrain the effective temperature, the results have been combined
into a single abundance estimate for each star. For our hottest target,
SK\,213, the results are affected by the choice of microturbulence but in
general the estimates are compatible with other analyses of SMC objects.

The principle conclusion from the model atmosphere analysis is that  the SMC
Wing targets appear to have similar chemical compositions to those of other
objects in the SMC. In particular there is no evidence for the low metallicity
found in the ICR targets by Rolleston et al.\ (\cite{Rol99}). From
Fig.~\ref{smcw_fig2}, it can be seen that all four targets are close to the
inter-cloud region and it is arguable that two targets (ICR\,02134-7431 and
SK\,213) are indeed part of the ICR or at the boundary between the Wing and the
ICR. Given this close proximity, the difference in the chemical composition
found by Rolleston et al.\ and in the present study is surprising  especially
as the ICR is thought to have formed from material from the SMC and LMC. The
low metallicity found by Rolleston et al.\ implies that the most likely source
of the material is the SMC and then one is forced to the  conclusion that the
ICR originated from a region close to (or indeed part of) the Wing but with a
different chemical composition to that found for the parts of the Wing that we
have sampled. 

An alternative explanation is that the results in Rolleston et al.\ are
unreliable and that the ICR has a metallicity consistent with the SMC (or LMC).
The low metallicity of the ICR was based on an LTE analysis of three targets
with effective temperature in the range 20\,000 to 24\,000~K and logarithmic
gravities of 3.6 to 3.8 dex; this would correspond to a spectral classification
of early-B-type giants/dwarfs. The analysis was undertaken differentially with
respect to two galactic stars with very similar atmospheric parameters, whilst
the hottest ICR target (DI\,1239) was also analysed relative to the SMC main
sequence star AV\,304, which has an effective temperature approximately 
2\,500~K higher. The results for all three stars were consistent and indicated
that the metallicity of the ICR was between 1.0 and 1.2 dex lower than our
Galaxy for the elements C, N, O, Mg and Si.  Additionally the abundances of O,
Mg and Si were found to be 0.5 to 0.7 dex lower in DI\,1239 compared with
AV\,304. One of the principal areas of uncertainty in the Rolleston et al.\
analysis was the use an LTE approximation.

\begin{table}  

\caption{A non-LTE analysis of the ICR stars using TLUSTY. For each star, the
newly derived atmospheric parameters are followed by their absolute abudances
along with the differential abundances relative to AV\,304. Where more than one
lines were used in the analysis, the standard deviation of the sample is
$\le$ 0.3 dex.}\label{tab:icrs}

\begin{tabular}{clllllll}
\hline\hline\noalign{\smallskip}
         && \multicolumn{2}{c}{DI\,1162} & \multicolumn{2}{c}{DI\,1239} & \multicolumn{2}{c}{DGIK\,975}\\
\noalign{\smallskip}\hline\noalign{\smallskip}
 \teff\/ [K]    &&  21\,500  &           & ~~\,24\,500  &               & ~~\,21\,000  &               \\
 log~$g$ [cm]   &&  3.25     &           & ~~\,3.70	&               & ~~\,3.35     &               \\
 \vt\/   [\kms] &&  5        &           & ~~\,5	&               & ~~\,5	       &               \\
\noalign{\smallskip}
C\,{\sc ii}     &&  7.2 ( 2) & --0.3 ( 2)& ~~\,7.5      & ~~\,+0.2	& $<$ 6.8 ( 2) & $<$ --0.6 ( 2)\\
N\,{\sc ii}     &&  6.5      & --0.1     & $<$7.3 ( 2) & $<$ +0.5      & ~~\,6.5      & ~~\,--0.1     \\
O\,{\sc ii}     &&  7.8 ( 6) & --0.3 ( 6)& ~~\,7.6 (11) & ~~\,--0.5 (10)& ~~\,7.9 (10) & ~~\,--0.2 (10)\\
Mg\,{\sc ii}    &&  6.3      & --0.5     & ~~\,6.3      & ~~\,--0.4	& ~~\,6.0      & ~~\,--0.8     \\
Si\,{\sc ii/iii}&&  6.5 ( 5) & --0.5 ( 3)& ~~\,6.0 ( 3) & ~~\,--0.8 ( 3)& ~~\,6.2 ( 5) & ~~\,--0.9 ( 3)\\
{\bf Mean$^*$}  &&           &{\bf $-$0.4 $\pm$ 0.2} & & {\bf $-$0.4 $\pm$ 0.4}&&
{\bf $-$0.6 $\pm$ 0.3} \\  
\noalign{\smallskip}\hline\noalign{\smallskip}
\end{tabular}\\
\smallskip * Excluing N.
\end{table}

To investigate this, we have re-analysed the ICR stars in Rolleston et al.\
(1999) using the same non-LTE grids used in this study. The atmospheric
parameters were re-deduced using the same methodology adopted by Rolleston et
al., i.e.\  the effective temperature was derived using the Balmer discontinuity
and the surface gravity from the comparison of the observed Balmer line profiles
with the non-LTE theoretical spectra.  Using these new atmospheric parameters,
we have derived the absolute non-LTE abundance of the ICR stars and detailed
line-by-line differential abundances relative to AV\,304 (summary in 
Table~\ref{tab:icrs}).  While the non-LTE atmospheic parameters are different
from their LTE counterparts, the derived absolute abundances agree very well
with the corresponding LTE values within the error. So do the differential
abudances relative to the SMC standard AV\,304.  More importantly, the metal
deficiency seen in Rolleston et al.\ is still evident in our non-LTE analysis.
With an exception of C and N for which the ICR stars show both over- and
under-abundances, other metal lines (O, Mg and Si) all show similar amount of
under-abundace as seen in Rolleston et al.  Rolleston et al.\ reported the ICR
stars have metallicity $\sim -$0.5 dex lower  than that of the SMC.  Excluding
N, which might have been contamined from the dredged-up material, the average of
our differential analysis gives a similar value. 

The other main uncertainty is probably the choice of the microturbence. This
quantity is difficult to estimate and its choice would affect the stronger metal
lines in the Galactic stellar spectra more than those for the ICR targets.
Hence, for example, adoption of larger microturbulences would reduce the
magnitude of the underabundance found in the ICR. However, the microturbulences
found by Rolleston et al.\ from the \ion{O}{ii} spectra range from 5 to 10 km
s$^{-1}$ and appear reasonable for the stellar atmospheric parameters (see, for
example, Hardorp and Scholz \cite{Har70}; Gies \& Lambert \cite{Gie92}; Kilian
et al.\ \cite{Kilian}). 

Lehner et al.\ (\cite{Leh01}) has studied the interstellar ultraviolet
absorption and \ion{H}{i} emission spectrum towards a young star in Magellanic
Bridge. These observations also appeared to support a low metallicity ($\simeq
-1.1$ dex) compared to that of the  Galaxy. Hence, although the failure to
identify a low metallicity component in the SMC Wing is worrying, there would
appear to be no substantial reason to discount the results of Rolleston et al.\ 
(\cite{Rol99}). Clearly further observations of targets in both the ICR and the
SMC/ICR interface are crucial to understanding this discrepancy.

\begin{acknowledgements} 

We are grateful to the staff of the Anglo-Australian Telescope for their
assistance during the observing runs. JKL and WRJR acknowledge financial
support of the PPARC. We are grateful to Ivan Hubeny and Thierry Lanz for their
help in running the non-LTE code {\sc tlusty} and to Ian Hunter for making
available preliminary results for his non-LTE analysis of AV\,304. We thank an
anonymous referee for encouraging us to re-analyse the ICR targets using our
non-LTE grid.

\end{acknowledgements}

\end{document}